\documentclass[a4paper,fleqn,usenatbib]{mnras}

\usepackage{newtxtext,newtxmath}

\usepackage[T1]{fontenc}
\usepackage{ae,aecompl}
\def\mdot{$\dot{M}$}

\usepackage{graphicx}	
\usepackage{amsmath}	
\usepackage{amssymb}	
\usepackage{array}

\title[Evolution of RSG mass-loss rates]{The evolution of Red Supergiant mass-loss rates}

\author[E. R. Beasor \& B. Davies]{
Emma R. Beasor$^{1}$\thanks{E-mail:e.beasor@2010.ljmu.ac.uk}
\& Ben Davies$^{1}$
\\
$^{1}$Astrophysics Research Institute, Liverpool John Moores University, Liverpool,  L3 5RF, UK}

\date{Accepted XXX. Received YYY; in original form ZZZ}

\pubyear{2017}

\begin{document}
\label{firstpage}
\pagerange{\pageref{firstpage}--\pageref{lastpage}}
\maketitle

\begin{abstract}
The fate of massive stars with initial masses >8M$_\odot$ depends largely on the mass-loss rate (\mdot ) in the end stages of their lives. Red supergiants (RSGs) are the direct progenitors to Type II-P core collapse supernovae (SN), but there is uncertainty regarding the scale and impact of any mass-loss during this phase. Here we used near and mid-IR photometry and the radiative transfer code DUSTY to determine luminosity and \mdot\ values for the RSGs in two Galactic clusters (NGC 7419 and $\chi$ Per) where the RSGs are all of similar initial mass ($M_{\rm initial}$$\sim$16M$_\odot$), allowing us to study how \mdot\ changes with time along an evolutionary sequence. We find a clear, tight correlation between luminosity and \mdot\ suggesting the scatter seen in studies of field stars is caused by stars of similar luminosity being of different initial masses.  From our results we estimate how much mass a 16M$_\odot$ star would lose during the RSG phase, finding a star of this mass would lose a total of 0.61$^{+0.92}_{-0.31}$M$_\odot$. This is much less than expected for \mdot\ prescriptions currently used in evolutionary models.  
\end{abstract}




\section{Introduction}

Single stars with initial masses between 8 - 25 M$_\odot$ are predicted to evolve to become red supergiants (RSGs) before they end their lives as core collapse supernovae (SNe). During this phase, the stars become extremely luminous and undergo strong mass-loss. The driving mechanism for these winds remains uncertain and so mass-loss rates (\mdot) cannot be calculated from first principles, and instead requires observations to provide input to stellar evolution models. 

Mass-loss during the RSG phase can have a significant effect on the subsequent evolution of the star. If a large amount of mass is lost, the star may evolve back to the blue of the Hertzsprung-Russel diagram (HRD) rather than remaining on the RSG branch and exploding as a Type II-P SN \citep{georgy2015mass}. Observational studies have identified an apparent lack of high mass Type II-P SN progenitors \citep[>17M$_\odot$,][]{smartt2009death,smartt2015observational} with enhanced \mdot\ during the RSG phase being suggested as a potential solution to this. For example, the progenitor to Type IIb SN 2011dh has been identified to be a 13M$_\odot$ yellow supergiant \citep[YSG,][]{maund2011yellow}. Single star evolutionary models predict that stars of this mass will become RSGs and explode as Type II-P SN. However \cite{georgy2012yellow} showed that when \mdot\ is increased during the RSG phase by 10-15 times, the stars evolve to higher effective temperatures and explode as progenitors more comparable to YSGs. Therefore, a potential explanation for the `missing' high mass progenitors is that they evolve away from the RSG phase before explosion due to high mass loss, possibly exploding as a different kind of SN \citep{smith2011observed}.       

All evolutionary models use empirically derived \mdot-prescriptions for RSGs when calculating the evolution of massive stars. These prescriptions are taken from studies of large numbers of field stars, where \mdot\ is measured for stars of varying luminosity. From this, an empirical relation between \mdot\ and luminosity can be found. There are a number of prescriptions available \citep[e.g.][]{van2005empirical, reimers1975circumstellar,nieuwenhuijzen1990parametrization,feast1992ch}, with the most commonly used prescription being that of de Jager \citep[][hereafter DJ88]{de1988mass}. Large systematic offsets exist between these prescriptions, often by a factor of 10 or more \citep[see Figure 1 in ][]{mauron2011mass} and there is also large internal scatter within each prescription again by up to a factor of 10 \citep[e.g. see Figure 5 in ][]{mauron2011mass}. Therefore, a given evolutionary model will adopt a different \mdot\ at a given luminosity depending on which prescription is in use. 

Recent work has shown that this dispersion is reduced when looking at RSGs in a cluster \citep[][hereafter BD16]{beasor2016evolution}, where all of the RSGs can be assumed to have the same metallicity, the same age and similar initial mass.  In BD16, \mdot\ and luminosities were derived for RSGs in NGC 2100, a cluster in the Large Magellanic Cloud (LMC), finding a tight correlation between \mdot\ and luminosity with little scatter. This suggests that the origin for the dispersion in previous \mdot\ studies comes from differences in initial masses of the stars, their metallicities, or a combination of the two. BD16 also showed that \mdot\ increases as  a star evolves and found little justification for increasing \mdot\ by more than a factor of 2 during the RSG phase, as suggested by \cite{georgy2012yellow}. 

We now present a similar study, this time focussing on two Galactic clusters, NGC 7419 and $\chi$ Per. Both clusters contain RSGs at different evolutionary stages all of a similar initial mass and Solar metallicity.  Using near and mid-IR photometry we have derived \mdot s and luminosities for 13 RSGs, allowing us to study how \mdot\ changes with evolution at a fixed metallicity. In Section 2 we discuss our modelling procedure and justifications for the parameters chosen. In Section 3 we discuss the application of our fitting methodology to Galactic clusters NGC 7419 and $\chi$ Per and the results we derive. In Section 4 we compare our results with commonly used \mdot\ prescriptions, and calculate the total mass lost for an RSG of a given initial mass during the RSG phase. In Section 5 we present our conclusions.

\section{Application to Galactic clusters}
\subsection{Sample selection}
In this paper we have chosen to study the Galactic clusters NGC 7419 and $\chi$ Per (also known as NGC 884), both of which contain a number of RSGs at Solar metallicity. These clusters have been found to be of similar ages \citep[$\sim$ 14Myr,][]{marco2013ngc,currie2010stellar}, which means all of the RSGs within each cluster have comparable initial masses (in this case 16M$_\odot$, see Section 2.2). As the RSG phase is short \citep[$\sim$10$^6$yrs, ][]{georgy2013grids} we can assume the stars are all coeval, i.e. any spread in age between the stars is small compared to the lifetime of the cluster. A coeval set of RSGs also allows us to use luminosity as a proxy for evolution, since those stars with higher luminosities have evolved slightly further up the RSG branch. 

The photometry used in this work is shown in Table 1 and is taken from 2MASS, WISE and MSX \citep{skrutskie2006two,wright2010wide,price2001midcourse}. The stars selected were known cluster members. For $\chi$ Per we picked RSGs within 6' of cluster centre, which is the distance to the edge of the $h$ \& $\chi$ Per complex, to maximise the probability that the stars were cluster members and hence were formed at the same time. However, \cite{currie2010stellar} showed that everything within the h $\&$ $\chi$ Per complex, including the surrounding region, is the same age to within the errors.

\subsection{Initial masses}
To estimate initial masses for the RSGs, we need to know the age of the cluster. We have taken the best fit isochrone for both clusters \citep[$\sim$ 14Myr from Padova isochrones,][]{marco2013ngc,currie2010stellar} as well as Geneva rotating and non-rotating isochrones. We compare the best fit turn off mass to that of other evolutionary models to determine a model dependent age for each cluster, and therefore the model dependent mass for the RSGs. From this, we are also able to ensure that we are comparing a self consistent age and mass for each evolutionary model.

From the original Padova isochrone, a turn-off mass of 14M$_\odot$ is found and an RSG mass of $\sim$ 14.5M$_\odot$ for both clusters. The non-rotating Geneva models suggest the cluster's turn-off mass is best fit by a 10Myr isochrone, giving an RSG mass of 17-18M$_\odot$.  The rotating models suggest an age of 14Myrs, with an RSG mass of  15-16M$_\odot$. For the rest of this paper we will assume the initial mass for the stars across both MW clusters is 16M$_\odot$, in between the rotating and non-rotating estimates\footnote{The evolutionary models suggest that for a single age cluster (e.g. Geneva rotating, 14Myrs) the difference in initial mass between stars at the start of the RSG phase and stars at the end of the RSG phase is $\sim$0.8M$_\odot$. A significant dispersion in initial masses between the RSGs in our sample is therefore unlikely. }.

\setlength{\extrarowheight}{6pt}
\begin{table*}
\centering
\tiny
\caption{Observational data for RSGs in $\chi$ Per \& NGC 7419. All fluxes are in units of Jy. All photometry for WISE 1 and 2 are upper limits. }
\label{my-label}
\begin{tabular}{lcccccccccccccccc}
\hline\hline
 Name  & 2MASS-J & 2MASS-H & 2MASS-Ks & WISE1& WISE2 & WISE3 & WISE4  & MSX-A & MSX-C & MSX-D & MSX-E \\
 & & & & (3.4  $\micron$)& (4.6  $\micron$)& (11.6  $\micron$) &(22  $\micron$) \\
\hline
 FZ Per& 48.18$\pm$  3.22& 70.32$\pm$  4.84& 67.79$\pm $ 6.56&$   - $&$< $64.90&   9.33$\pm $  0.06&  4.43$\pm$  0.06& 12.20& 10.90&  6.52&  4.13  \\ 
 RS Per& 95.69$\pm$  6.83&146.93$\pm$ 13.38&158.20$\pm$ 21.27&$<$ 952.05&$<$317.56&  51.70$\pm$  39.43& 43.88$\pm$  0.07& 57.60& 59.50& 42.80&41.10 \\ 
 AD Per&$ 70.94\pm  4.95$&$104.30\pm  8.66$&$111.38\pm 14.90$&$< 349.83$&$<115.72$&$  18.09\pm   2.05$&$ 11.16\pm  0.36$& 20.80& 20.40& 13.10& 11.30 \\ 
V439 Per&$ 35.52\pm  2.46$&$ 52.71\pm  3.80$&$ 55.98\pm  6.24$&$< 119.42$&$< 45.69$&$   4.29\pm   0.05$&$  1.85\pm  0.02$&  7.33&  5.18&  3.20&  2.39 \\ 
V403 Per&$ 25.85\pm  0.00$&$ 42.22\pm  0.00$&$ 42.66\pm 14.29$&$<   0.23$&$<  0.02$&$   2.89\pm   0.02$&$  0.93\pm  0.01$&  5.31&  2.71& 1.83& -1.86 \\ 
V441 Per&$ 65.12\pm  4.84$&$105.66\pm  9.43$&$101.76\pm 11.87$&$< 511.75$&$<165.28$&$  16.54\pm   1.17$&$ 11.55\pm  0.47$& 19.30& 16.80& 13.30& 11.10 \\ 
 SU Per&$118.27\pm 10.64$&$173.42\pm 16.55$&$174.58\pm 26.64$&$<1099.16$&$<315.23$&$  39.14\pm  25.30$&$ 27.13\pm  0.11$& 43.80& 40.00& 24.10& 30.10 \\ 
 BU Per&$ 53.91\pm  4.28$&$ 86.84\pm  6.61$&$ 88.39\pm  9.27$&$< 467.15$&$<170.70$&$  32.05\pm1057.73$&$ 28.15\pm  0.11$& 33.10& 36.70& 26.30& 30.20 \\ 
\hline
 MY Cep&$ 23.40\pm  1.43$&$ 65.81\pm  5.34$&$ 93.07\pm 12.10$&$< 712.95$&$<239.12$&$  76.04\pm  28.26$&$ 81.94\pm  0.03$& 87.80&134.00& 97.00& 97.70 \\ 
  BMD 139 &$  6.00\pm  0.02$&$ 16.02\pm  0.13$&$ 17.75\pm  0.16$&$<   0.56$&$<  0.49$&$   0.10\pm   0.00$&$  0.12\pm  0.00$&  3.39&  2.41&  1.58& -2.45 \\ 
  BMD 921&$  5.54\pm  0.02$&$ 12.32\pm  0.08$&$ 13.50\pm  0.05$&$<   9.45$&$<  5.64$&$   0.91\pm   0.00$&$  0.39\pm  0.00$&  1.71&  0.71&  0.74& -2.23 \\ 
  BMD 696&$  8.17\pm  0.03$&$ 13.98\pm  0.64$&$ 20.61\pm  0.20$&$<  31.01$&$<  9.35$&$   2.12\pm   0.01$&$  1.03\pm  0.01$&  3.40&  2.00&  1.68& -2.23 \\ 
  BMD 435&$  8.88\pm  0.03$&$ 14.99\pm  0.75$&$ 16.80\pm  1.19$&$<   0.60$&$<  1.12$&$   0.26\pm   0.00$&$  0.27\pm  0.00$&  3.43&  2.11&  1.33&  3.62 \\ 
\end{tabular}
\end{table*}

\section{Dust shell models}
The dust shell models used in this project were made using DUSTY \citep{ivezic1999dusty} which solves the radiative transfer equation for a central star surrounded by a spherical dust shell of a certain optical depth ($\tau_V$, optical depth at 0.55$\micron$), inner dust temperature ($T_{\rm in}$) at the inner most radius ($R_{\rm in}$) and radial density profile ($\rho_r$). Below we briefly describe our choices for the model input parameters and our fitting methodology, for an in depth discussion see \cite{beasor2016evolution}.  

\subsection{Model Setup}
The dust layer surrounding RSGs absorbs and reprocesses the light emitted from the star, with different compositions of dust affecting the spectral energy distribution (SED) in different ways. We have opted for oxygen rich dust as specified by \cite{draine1984optical} and a grain size of 0.3$\micron$ \citep[e.g.][]{smith2001asymmetric,scicluna2015large}. 

To calculate mass-loss rates we have assumed a steady state density distribution falling off as $r^{-2}$. Departure from this law has been suggested for some RSGs \citep[e.g.][]{shenoy2016searching}, a matter which we discuss in detail in \cite{beasor2016evolution}. As we do not have outflow velocity measurements for the RSGs in our sample, we have assumed a uniform speed of 25$\pm$5 km s$^{-1}$, consistent with previous measurements \citep[e.g.][]{van2001circumstellar, richards1998maser}. 

We have also assumed a gas-to-dust ratio ($r_{\rm gd}$) of 200 and a grain bulk density ($\rho_{\rm d}$) of 3 g cm${^-3}$. From this, we can then calculate \mdot\ values from the following equation
 \begin{equation} \dot{M} = \frac{16\pi}{3} \frac{R_{in} \tau_{V}  \rho_d a v_\infty}{Q_{V}}r_{gd}
 \end{equation}

\noindent where $Q_V$ is the extinction efficiency of the dust \citep[as defined by the dust grain composition,][]{draine1984optical}.

The stellar effective temperature $T_{\rm eff}$ changes the position of the peak wavelength of the SED. For NGC 884,  the RSGs are of spectral types M0 -  M3.5, corresponding to an approximate temperature range of 3600K - 4000K \citep[taken from the temperature scale of ][]{levesque2005physical}. In contrast \cite{gazak2014quantitative} found a narrower $T_{\rm eff}$ spread among the stars in $\chi$ Per, 3720K - 4040K. In this work, we have opted for a fiducial SED of 3900K for the analysis of this cluster, with the errors on $L_{\rm bol}$ found by rerunning the analysis with SEDs of temperatures $\pm$ 300K fully encompassing the observed range of both \cite{gazak2014quantitative} and \cite{levesque2005physical}.

For the NGC 7419 RSGs,  the spectral types range from M0 to M7.5 \citep{marco2013ngc} corresponding to a temperature range of 3400 - 3800K. For this cluster, we chose to use a fiducial SED of 3600K with further analysis completed using SEDs at 3400K and 3800K. 

To ensure the robustness of out $T_{\rm eff}$ assumptions we also systematically altered the fiducial value for each cluster and re-derived luminosities and \mdot. By doing this we found that altering the $T_{\rm eff}$ by $\pm$300K caused the value of \mdot\ to change by $\pm$5\%, while luminosity was only affected by around 0.1 dex. 

In this study, we have again allowed $T_{\rm in}$ and $\tau_V$ to be free parameters to be optimised by the fitting procedure. $T_{\rm in}$ defines the temperature of the inner dust shell (and hence its position, $R_{\rm in}$) while optical depth determines the dust shell mass. The fitting methodology is described in the next subsection. 
\subsection{Fitting methodology}
We computed two grids of dust shell models for each SED spanning a range of inner temperatures and optical depths. The first grid spanned $\tau_V$ values of 0 - 1.3, while the second grid spanned $\tau_V$ values of 0 - 4, each having 50 grid points, and each having $T_{\rm in}$ values of 0 - 1200K in steps of 100K\footnote{For MY Cep, as the $\tau_V$ range in our initial model grid was not high enough to match the observed photometry, we had to use a coarser model grid with a large range of $\tau_V$ values.}. For each model output spectrum, we created synthetic photometry by convolving the model spectrum with the relevant filter profile. By using ${\chi^2}$ minimisation we determined the best fitting model to the sample SED. 
\begin{equation}
\chi^2 =\sum_i \frac{ (O_{i}-E_{i})^2 }{\sigma_i^2}
\end{equation}
where $O$ is the observed photometry, $E$ is the model photometry, $\sigma$$^2$ is the error and $i$ denotes the filter. In this case, the model photometry provides the ``expected'' data points. The best fitting model is that which produced the lowest  ${\chi^2}$. 

Some of the photometric points used in this study were upper limits, and therefore these data were used to preclude models for which the synthetic photometry exceeded these limits. As well as this, any photometric point that had an error of <10\% had a blanket error of 10\% applied to account to systematic errors. The errors on our fitting results are defined as the minimum $\chi^2$ value + 10, defined to allow stars with the lowest measured \mdot\ values which were consistent with non-detections to have \mdot\ values that are upper limits only. 

\section{Modelling results}
\begin{figure*}
  \caption{ \textit{Left panel:}  Model plot for SU Per including all contributions to spectrum. The ``error models" are the models that fit within the minimum $\chi^2$+10 limit. The silicate bump at 10$\micron$ is clearly visible on the spectra suggesting a large amount of circumstellar material. \textit{Right panel:} Contour plot showing the degeneracy between $\chi^2$ values and best fitting \mdot\ values in units of 10$^{-6}$ M$_\odot$ yr$^{-1}$. The red contour highlights the models within the minimum $\chi^2$+10 limit. }
  \centering
  \label{fig:allcont}
     \includegraphics[height=7cm]{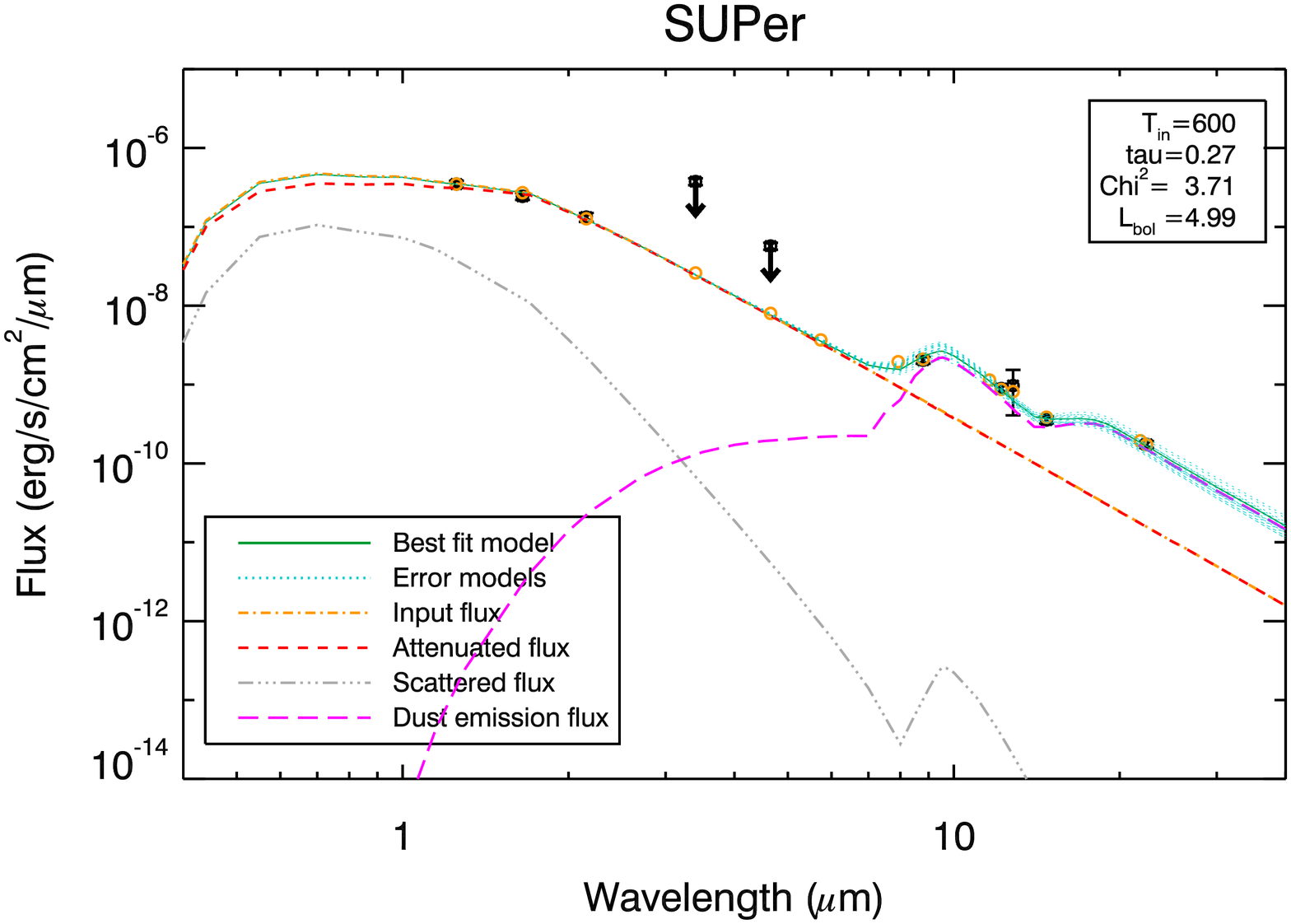}
     \includegraphics[height=7cm]{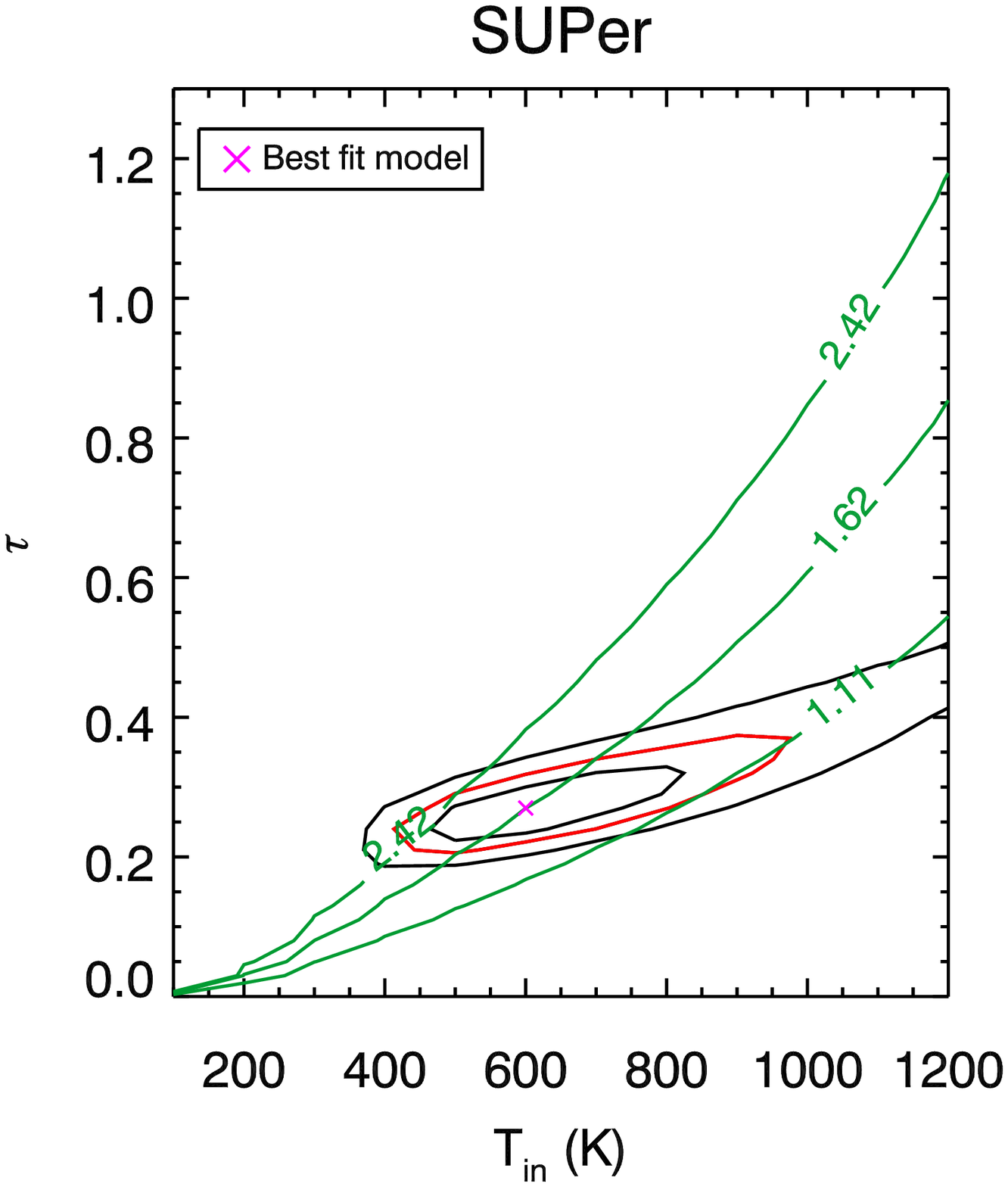}
\end{figure*}

We ran our fitting procedure for all of the RSGs in our sample. Our results for $\chi$ Per and NGC 7419 are shown in Table 2. 
Figure \ref{fig:allcont} shows the best fit model for the brightest star in these clusters, SU Per, with all contributions to the output spectrum. The left panel of the plot shows the best fit model spectra (green line), the models within our error range (blue dotted lines) as well as the photometric points, where the black crosses shown the real photometry and orange circles show the model photometry. The right hand panel shows our best fit model located on a $T_{\rm in}$ - $\tau$ plane with the mass loss rate isocontours overplotted. 

Both clusters are affected by foreground reddening. To correct for this we used the published extinction laws for the 2MASS, MSX and WISE photometry \citep[][]{koornneef1983near, messineo200586, gontcharov2016extinction}. We adopted foreground V-band extinctions of 1.66 and 5.27 for the clusters $\chi$ Per and NGC 7419 respectively \citep{currie2010stellar,marco2013ngc}. There is evidence that differential extinction is present in each cluster. For $\chi$ Per, \cite{currie2010stellar} find a V-band dispersion of 0.09 mag (estimated from J-K colours), equivalent to $\sim$0.01 mag in $K_{\rm s}$. This level of differential extinction is smaller than the errors on our photometry and hence will not effect our modelling results. Similarly, the differential reddening across NGC 7419 is approximately 0.2 mag in $K_{\rm s}$ \citep[][where individual reddenings were calculated for all cluster members]{marco2013ngc}, comparable to the photometric error.

The results of our modelling  for all stars in the clusters are shown in Table 2. The luminosities are bolometric as found in \citep{davies2017initial} where possible, else they are calculated by integrating under the best fit spectra with errors on $L_{\rm bol}$ dominated by the uncertainty in $T_{\rm eff}$. It can be seen that the stars with the highest mass loss rates have inner dust temperatures that are more constrained, while lower \mdot\ stars show a larger spread in $T_{\rm in}$. We also find that the best fits were achieved when allowing $T_{\rm in}$ to vary from the dust sublimation temperature of 1200K. The the stars with the highest \mdot\, $T_{\rm in}$ is typically around 600K rather than the canonical 1200K dust sublimation temperature. For B921, as we could only place an upper limit on the optical depth it was not possible to constrain an inner dust temperature, and hence we have plotted an upper limit for the value of \mdot. 

When plotting $L_{\rm bol}$ versus \mdot\ a clear positive correlation can be seen (see Fig. \ref{fig:mdotcomp}), demonstrating an increasing \mdot\ with evolution. We also include results from NGC 2100 (BD16). Overplotted are also some commonly used \mdot\ prescriptions (assuming a $T_{\rm eff}$ of 3900K), including; \cite{de1988mass}, \cite{reimers1975circumstellar}, \cite{van2005empirical} and \cite{goldman2017wind}.

\begin{table}
\centering
\tiny
\caption{Fitting results for the RSGs in $\chi$ Per and NGC 7419.  Bolometric luminosities are from \citep{davies2017initial}.}
\label{my-label}
\begin{tabular}{lcccccc}

\hline\hline
Star & $T_{\rm in}$ (K) & $\tau_V$ &  \mdot\ (10$^{-6}$M$_\odot$ yr$^{-1}$) &  $L_{\rm bol}$   \\ [0.5ex] 
\hline
 FZ Per&$1000^{+ 200}_{- 400}$&$0.19^{+0.08}_{-0.06}$&$ 0.30^{+ 0.18}_{- 0.07}$&$ 4.64^{+ 0.06}_{- 0.05}$ \\
 RS Per&$ 600^{+ 200}_{- 200}$&$0.53^{+0.13}_{-0.08}$&$ 3.03^{+ 2.31}_{- 0.94}$&$ 4.92^{+ 0.18}_{- 0.07}$ \\
 AD Per&$ 600^{+ 600}_{- 100}$&$0.21^{+0.16}_{-0.02}$&$ 0.97^{+ 0.33}_{- 0.50}$&$4.80^{+ 0.08}_{- 0.05}$ \\
V439 Per&$1200^{+   0}_{- 500}$&$0.11^{+0.02}_{-0.03}$&$ 0.10^{+ 0.10}_{- 0.01}$&$ 4.53^{+ 0.06}_{- 0.05}$ \\
V403 Per&$1200^{+   0}_{- 400}$&$0.08^{+0.00}_{-0.03}$&$ 0.06^{+ 0.02}_{- 0.02}$&$ 4.41^{+ 0.06}_{- 0.05}$ \\
V441 Per&$ 600^{+ 300}_{- 200}$&$0.21^{+0.08}_{-0.02}$&$ 0.93^{+ 0.72}_{- 0.31}$&$ 4.75^{+ 0.10}_{- 0.06}$ \\
 SU Per&$ 600^{+ 300}_{- 100}$&$0.27^{+0.10}_{-0.06}$&$ 1.62^{+ 0.72}_{- 0.63}$&$ 4.99^{+ 0.09}_{- 0.05}$ \\
 BU Per&$ 500^{+ 200}_{- 100}$&$0.56^{+0.10}_{-0.08}$&$ 3.24^{+ 1.53}_{- 1.28}$&$ 4.67^{+ 0.07}_{- 0.05}$ \\

\hline

 MY Cep&$ 600^{+ 200}_{- 100}$&$2.04^{+0.41}_{-0.33}$&$18.04^{+ 7.15}_{- 8.54}$&$ 5.19\pm 0.07$ \\
  BMD 139&$ 900^{+ 300}_{- 400}$&$0.16^{+0.08}_{-0.00}$&$ 0.27^{+ 0.44}_{- 0.05}$&$ 4.55\pm 0.08$ \\
  BMD 921=56&$ - $&$<0.03 $&$< 0.06 $&$ 4.45\pm 0.10$ \\
  BMD 696=122&$ 700^{+ 500}_{- 200}$&$0.08^{+0.08}_{-0.00}$&$ 0.22^{+ 0.17}_{- 0.04}$&$ 4.63\pm 0.08$ \\
  BMD 435&$1100^{+ 100}_{- 300}$&$0.16^{+0.08}_{-0.00}$&$ 0.18^{+ 0.15}_{- 0.04}$&$ 4.54\pm 0.11$ \\

\end{tabular}
\end{table}

\section{Discussion}
\subsection{The \mdot\ - Luminosity Relation}
There are many empirical studies of RSG mass loss \citep[e.g.][]{reimers1975circumstellar,de1988mass,van2005empirical,de2010probing,bonanos2010spitzer,goldman2017wind} all showing significant scatter. For example the calibration of \cite{mauron2011mass} has a large peak-to-peak dispersion of a factor of $\sim$ 10. As these previous studies have focussed on field stars only, this internal scatter may be caused by inhomogeneity in the initial masses and/or metallicities of the stars in their samples (i.e. the stars are of different masses and ages at the same luminosity). 

\cite{goldman2017wind} studied RSG winds across a range of metallicities, stating that lower metallicity environments yield slower wind speeds for stars. As \mdot\ is directly proportional to the expansion velocity of the wind (see Equation 1), a lower wind speed will result in a lower derived \mdot. \cite{goldman2017wind} measure a relation between expansion velocity of the wind and metallicity ($v_{\rm exp}$$\propto$$ZL^{0.4}$), with derived expansion velocities then being compared to mass-loss rates. From this we can estimate how large the effect of varying metallicity is on \mdot. It is therefore possible that the $v_{\rm exp}$ we have assumed for the RSGs in NGC 2100 (an LMC metallicity cluster) is systematically high. However, we estimate this would reduce the \mdot\ values for these stars by around 25\%, bringing these results into even better agreement with the Galactic clusters. The effect of varying metallicity is therefore unlikely to be enough to cause the factor of 10 scatter seen in previous relations. 

\begin{figure*}
\centering
  \caption{Plot showing \mdot\ versus $L_{\rm bol}$ for all clusters we have studied. We also overplot our \mdot - luminosity relation for a 16M$_\odot$ star. }
  \centering
  \label{fig:mdotcomp}
    \includegraphics[width=\textwidth]{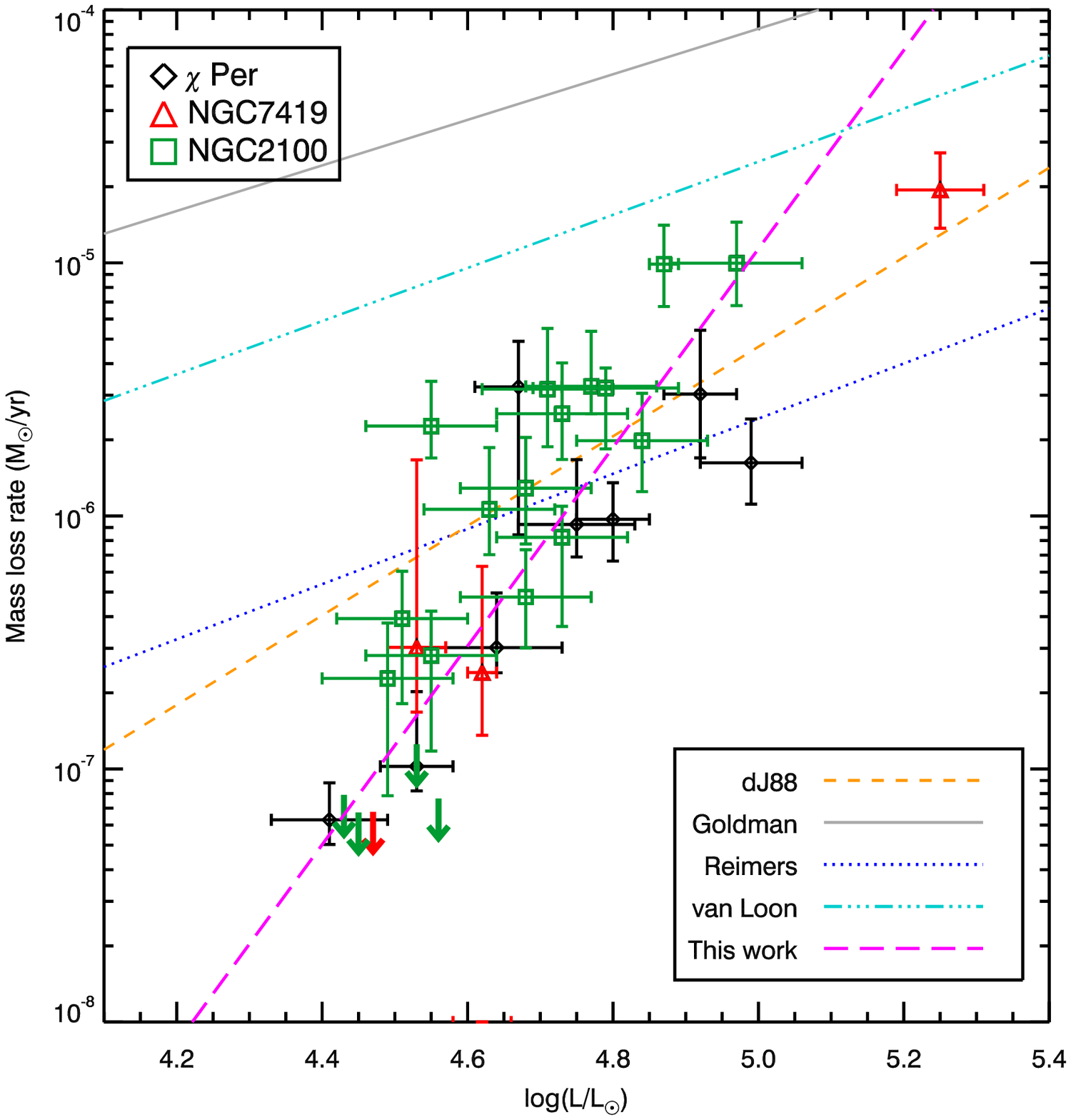}
\end{figure*}

The RSGs we have observed in NGC 2100, $\chi$ Per and NGC 7419 are all of a similar initial mass, but different metallicities (LMC, Solar and Solar, respectively). Despite this, there is still a tight correlation when all of the clusters are plotted together, Fig. \ref{fig:mdotcomp}, suggesting \mdot\ is only weakly depending on the metallicity\footnote{We have assumed a gas-to-dust ratios for each cluster, and we have also assumed the same expansion velocity for all clusters. This may be dependent on metallicity \citep{goldman2017wind}}. Though our data are within an order or magnitude of the DJ88 law, this law overestimated \mdot\ at low luminosities. The \cite{van2005empirical} and \cite{goldman2017wind} prescriptions both vastly overestimate \mdot\ at all luminosities compared to our results. This is likely due to both of these studies focussing on heavily dust enshrouded stars and/or maser emitters respectively, leaving their samples skewed to stars with the highest \mdot. These studies may be selecting stars near the end of their evolution, or with peculiar properties (for example, binarity). As the RSGs in our sample continue to evolve up the RSG branch, it is possible that they would eventually reach \mdot\ values as high as those observed for the \cite{goldman2017wind} and \cite{van2005empirical} samples. 

As all of the clusters we have looked at have been of a similar initial mass, we combine them to present a mass-loss rate prescription that will be applicable to stars of 16M$_\odot$ (Equation 3). As previously mentioned, our study suggests \mdot\ is only weakly dependent on metallicity between Solar and LMC and hence this relation depends only on the luminosity of the star. A linear best-fit to our data yield the relation,

\begin{equation}
\log(\dot{M} / M_\odot yr^{-1} )= a + b\log(L_{\rm bol} / L_{\odot})
\end{equation}
where a = -24.56 $\pm$ 1.65 and b = 3.92 $\pm$ 0.35, derived using IDL program FITEXY. We find a much steeper relation between mass loss and luminosity than previous studies, with a root mean square scatter of $\pm$0.4 dex. Our data suggest that when stars first join the RSG phase they have very low mass loss rates, implied by the small amount of circumstellar material present. The mass loss rate then increases by a factor of 100 throughout the RSG lifetime \footnote{Throughout this work we have assumed a constant gas-to-dust ratio for all stars in our sample. It is possible that this may change with evolution \citep[e.g.][]{mauron2011mass}, altering the \mdot\ - luminosity relation.}, while luminosity increases by only a factor of $\sim$5.

\subsection{The luminosity distribution of RSGs}
To determine the mass-loss rate at a given time-step, stellar evolution calculations use a mass-loss rate prescription (usually dJ88) in conjunction with the star's luminosity at that time step. If the luminosity is incorrectly estimated the adopted mass-loss rate during this phase will also be incorrect. 

We are able to test the implementation of \mdot\ in evolutionary models by looking at luminosity distributions. Luminosity distributions were generated by uniformly sampling RSG masses from a standard Salpeter IMF and assuming a uniform age for all stars. We simulated a large number of stars (10$^5$) within the relevant mass range and then normalised the distribution to match the total number of stars across both MW clusters. We also included the effect of measurement errors on the simulated luminosity distributions.

\begin{figure}
\centering
  \caption{Luminosity distributions for RSGs. Top panel: luminosity distribution for the 13 Galactic RSGs. Centre panel: luminosity distribution for RSGs in Geneva rotating models. Bottom panel: luminosity distribution for RSGs in Geneva non-rotating models. }
  \centering
  \label{fig:Lboldist}
    \includegraphics[width=\columnwidth]{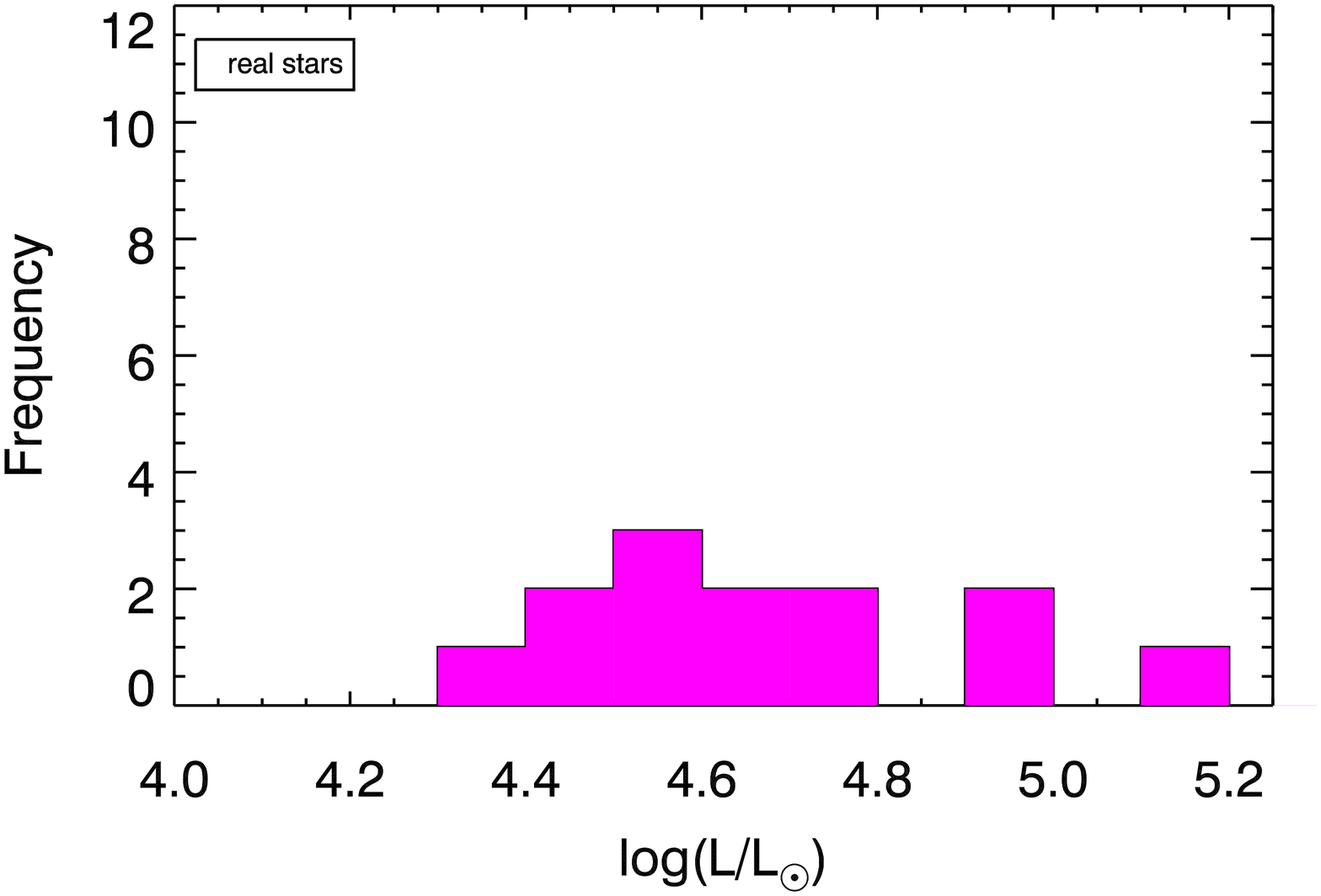}
    \includegraphics[width=\columnwidth]{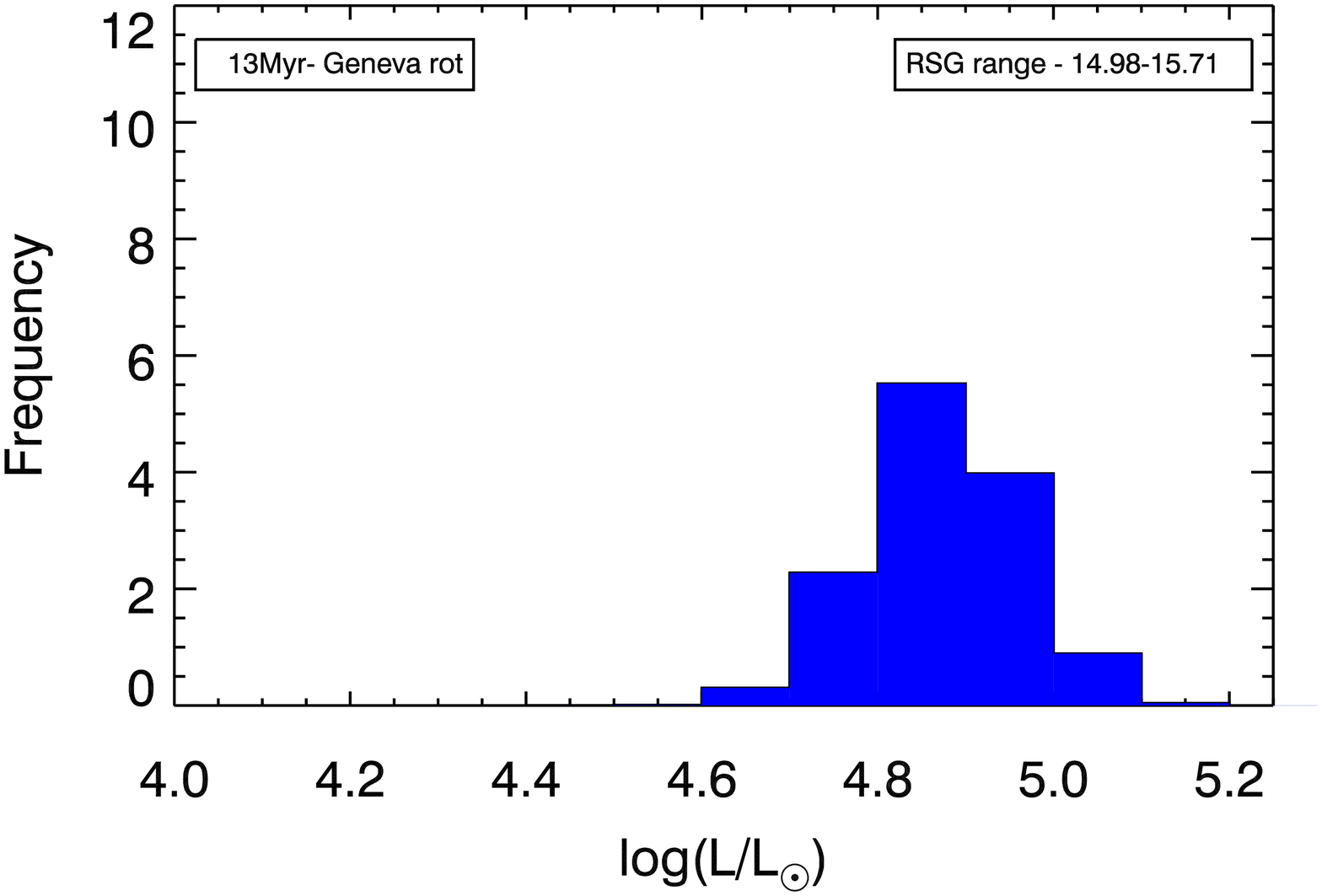}
    \includegraphics[width=\columnwidth]{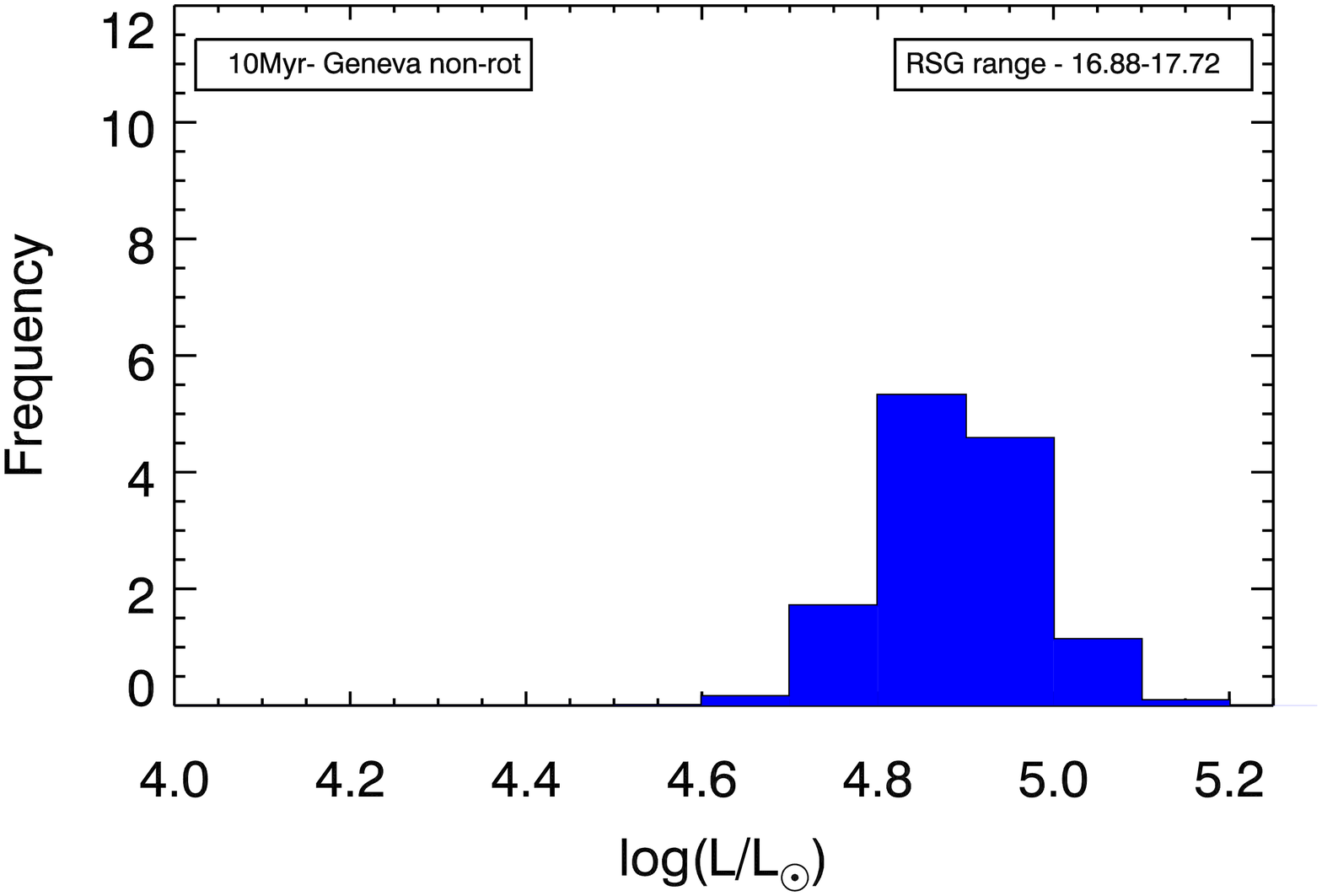}
\end{figure}

Figure \ref{fig:Lboldist} shows a comparison for the luminosity distribution of the RSGs\footnote{due to the difference in metallicity we have not included the NGC2100 stars in the luminosity distribution.} in the two MW clusters compared to the predicted luminosity distribution for 13 RSGs in the best fit Geneva non-rotating and Geneva rotating models (see Section 3.2). The observed luminosity distribution for the RSGs in the two clusters is peaked at a luminosity of log(L/L$_\odot$) $\sim$ 4.5, see the top panel of Fig. \ref{fig:Lboldist}. As the age spread between the RSGs in our sample is likely small, we can say that the different luminosities show the stars at slightly different stages of evolution. The concentration of stars in each luminosity bin indicates the relative amount of time the star spends at each interval. When comparing to the Geneva models, we see that RSGs are predicted by Geneva to spend most of their lives at much higher luminosities than observed, between log(L/L$_\odot$) $\sim$ 4.6 - 5.1, see Fig. \ref{fig:Lboldist}. As \mdot\ is scaled from luminosity in evolutionary models, this will lead to \mdot\ being overestimated by models throughout the RSG phase. 

As previously noted we have assumed a uniform age for all of the stars in the sample. However, given the errors on the age estimates for both clusters it is possible that an age spread exists between the RSGs. \cite{currie2010stellar} estimate the age error on $\chi$ Per to be $\pm$ 1Myr, while \cite{marco2013ngc} estimate the age error on NGC 7419 to be $\pm$ 2Myr. These numbers provide an upper limit to the age spread that exists within the clusters.

 We now investigate what effect a 2 Myr age spread would have on the observed luminosity distribution of the RSGs. We used Geneva isochrones and generated luminosity distributions for ages between 12 and 14 Myr (rotating models) at intervals of 0.1 Myr. We assumed a uniform age distribution, simulating a constant star formation rate for 2 Myr. This had no visible effect on the luminosity distribution, with the peak and width remaining the same. We could also have looked at what effect a Gaussian or exponentially declining star formation rate would have on the luminosity distribution, but as the constant rate would have the greatest impact it is unlikely either of these more complicated age spread functions would affect our conclusions. We can therefore rule out a non-instantaneous starburst as being the cause of the difference between the observed and predicted luminosity distributions. 

\subsubsection{Estimating the total mass lost during the RSG phase}
Having determined \mdot\ as a function of evolutionary phase, we now use our results to estimate the total mass lost as a 16M$_\odot$ star evolves up the RSG branch. This is important as the amount of mass lost can effect the appearance of the resulting SN. For example, while theory predicts RSGs to be the progenitors to Type II-P SN, there is the possibility that if the star were to lose a large enough mass it would appear instead as a Type IIn SN \citep[e.g.][]{smith2009red,smith2017endurance}. The narrow lines of a Type IIn SN are visible when a star explodes into a dense circumstellar medium, and for an RSG it would require $\sim$1M$_\odot$ material to be present around the star \citep{smith2009red}. There is increasing observational evidence for a continuum between II-L, IIn and II-P SN as opposed to the SN being produced by distinct progenitors. For example \cite{morozova2017unifying} found that II-L light curves could be fit by ordinary RSGs with dense CSM. Likewise, SN PTF11iqb showed narrow emission lines for it's first two days before they weakened and the light curve quickly began to resemble a II-L and II-P SN \citep{smith2015PTF11iqb}, which experienced enhanced mass-loss in the years preceding the SN.

We now investigate the total mass lost during the RSG phase for a typical 16M$_\odot$ star based on the RSGs in the two MW clusters. For a 16M$_\odot$ star, all major evolutionary codes agree that the lifetime of the RSG phase is 1Myr $\pm$ 15\% \citep{ekstrom2012grids,eldridge2009spectral,dotter2016mesa}. Using the luminosity distribution for the RSGs in $\chi$ Per and NGC 7419, we can deduce how much time an RSG spends at each evolutionary stage. As we know how \mdot\ varies with luminosity, it is possible to determine how \mdot\ varies with time. To convert luminosity into a time, we take the cumulative distribution of $L_{\rm bol}$ and interpolate this onto a time axis of 10$^6$ years. We then integrate \mdot\ with respect to time and estimate the total amount of mass that would be lost during the RSG phase for a 16M$_\odot$ star.

To estimate the error, we used a Monte-Carlo (MC) method. For each star studied, we randomly sampled its \mdot\ from an asymmetric Gaussian distribution centred on its best-fit value, with upper and lower 1-sigma widths determined by the upper and lower error bars. In each MC trial, we then integrated the \mdot\ of all stars with respect to time to find the total mass lost in that trial. By repeating $10^4$ times, we were able to determine the most likely total mass lost and the upper and lower 68\% confidence limits, comparable to a 1-sigma error bar. 

 From this we find a 16M$_\odot$ star would lose 0.61$^{+0.92}_{-0.31}$M$_\odot$ throughout the RSG phase. As this mass is lost over a long period of time (10$^{6}$yrs) it is unlikely there would be enough CSM close to the star to have an effect on the appearance of the resulting SN. Since the amount of envelope mass lost is small we can also expect there to be a long plateau in the SN light curve.

\begin{figure*}
\centering
  \caption{Plot showing the amount of mass lost for a star of a given initial mass for various stellar evolution models. The single pink circle shows the total amount of mass lost during the RSG phase for a 16M$_\odot$ star.}
  \centering
  \label{fig:masslost}
    \includegraphics[width=\textwidth]{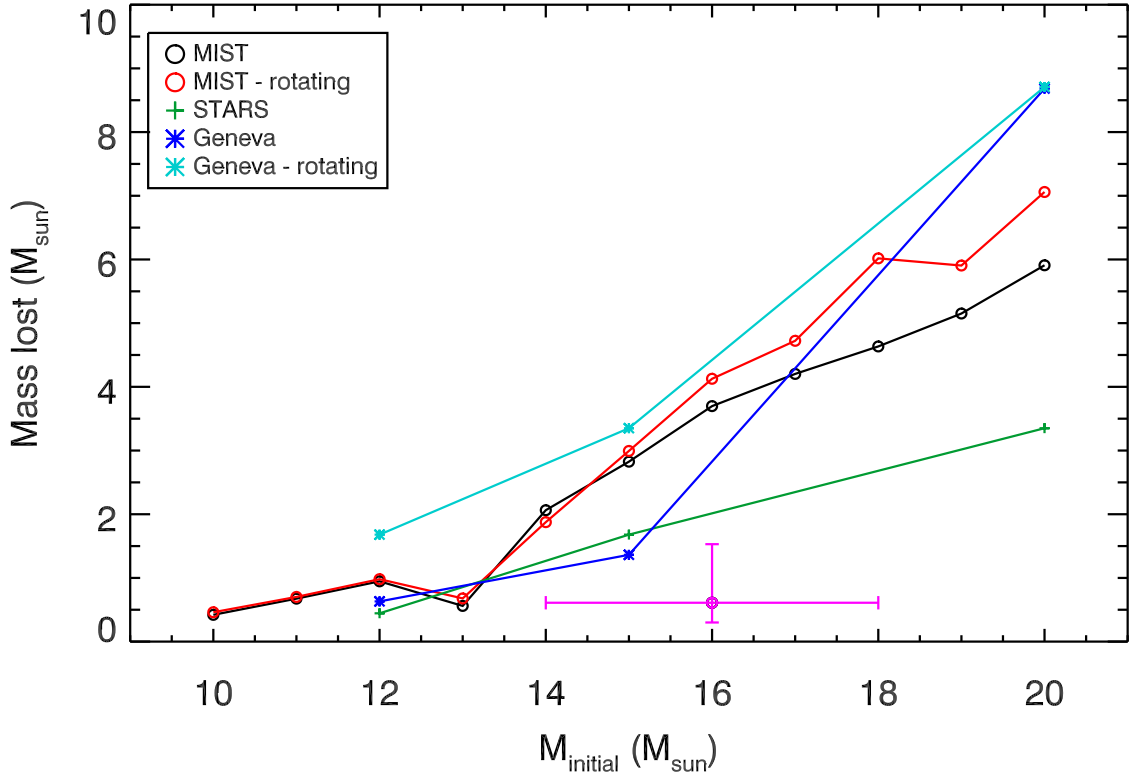}
\end{figure*}

We now compare our measurement of the total mass lost with predictions from stellar evolutionary models. Figure \ref{fig:masslost} shows how much mass is lost during the RSG phase as a function of initial mass for Geneva, STARS and MIST models \citep{ekstrom2012grids,eldridge2009spectral,dotter2016mesa}. This figure shows that compared to our observations for a 16M$_\odot$ star, all evolutionary models employ mass-loss rates that are too high, and hence over-predict the total mass lost during the RSG phase, with Geneva rotating models having the biggest offset from what we observe. In this figure we can also see that at $\sim$ 20M$_\odot$ the predicted total mass lost during the RSG phase from Geneva models deviates from the STARS model by around a factor of 3. This could be due to the Geneva group artificially enhancing \mdot\ by a factor of 3 for stars which exceed the Eddington limit in their evelopes by a factor of 5  \citep{ekstrom2012grids}. 

Figure \ref{fig:masslost} suggests that evolutionary models are in fact over predicting mass loss during the RSG phase, and increasing \mdot\ \citep[as suggested to solve the RSG problem, e.g.][]{georgy2012yellow} would only exacerbate this. Instead, a reappraisal of \mdot\ - prescriptions is needed to better inform stellar evolutionary models and allow more accurate predictions to be made. This is also a wider problem for massive star evolution, including hot star winds and luminous blue variables \citep[LBVs,][]{smith2014mass}. Future work will involve observing stars with different initial masses to fully understand the effect of \mdot\ throughout the RSG phase, and hence determine a new \mdot\ -luminosity relation. 

However, we have also shown that the luminosity distributions used in models are skewed to higher luminosities than observed in $\chi$ Per and NGC 7419. It may therefore be necessary to find a new way to implement \mdot\ in evolutionary models, since employing an empirical \mdot-luminosity relation will result in an \mdot\ which is too high if the models are over-predicting the RSG luminosities.

\section{Conclusions}
In this paper we have re-appraised the \mdot-luminosity relation using stars in young massive clusters. By focussing on RSGs within clusters we are effectively seeing the same star at different stages of evolution and can therefore observe how \mdot\ varies as the star evolves towards SN. We have determined \mdot's and luminosities for RSGs in Galactic clusters NGC 7419 and $\chi$ Per, both of which are approximately the same age and hence we are able to assume the RSGs are all of similar initial mass ($\sim$16M$_\odot$). From our study we can conclude the following:

\begin{enumerate}
  \item At fixed initial mass \mdot\ increases with time during the RSG evolution of the star, in a relation with little scatter. We suggest that the reason the correlation is tight compared to previous \mdot\ prescriptions is due to keeping $M_{\rm initial}$ constrained. We also find that this relation does not depend on the metallicity of the star, as we have studied both Galactic and LMC clusters and still find a tight correlation. From this we are able to present a new \mdot\ prescription for stars with initial masses of $\sim$ 16M$_\odot$ that depends only on the luminosity of the star.
  \item We have also compared the observed luminosity distribution for the RSGs in the two Galactic clusters to evolutionary models. We find that these models overpredict how much time the RSGs spend at high luminosities, and thus overpredict the total amount of mass-loss during the RSG phase. 
\end{enumerate}
\section*{Acknowledgements}
The authors would like to thank referee Eric Josselin for comments which helped improve the paper. We would also like to thank Nathan Smith for useful discussion and comments. We  acknowledge the use of the SIMBAD database, Aladin, IDL software packages and astrolib.





\bibliographystyle{mnras}
\bibliography{references} 

\begin{thebibliography}{}
\makeatletter
\relax
\def\mn@urlcharsother{\let\do\@makeother \do\$\do\&\do\#\do\^\do\_\do\%\do\~}
\def\mn@doi{\begingroup\mn@urlcharsother \@ifnextchar [ {\mn@doi@}
  {\mn@doi@[]}}
\def\mn@doi@[#1]#2{\def\@tempa{#1}\ifx\@tempa\@empty \href
  {http://dx.doi.org/#2} {doi:#2}\else \href {http://dx.doi.org/#2} {#1}\fi
  \endgroup}
\def\mn@eprint#1#2{\mn@eprint@#1:#2::\@nil}
\def\mn@eprint@arXiv#1{\href {http://arxiv.org/abs/#1} {{\tt arXiv:#1}}}
\def\mn@eprint@dblp#1{\href {http://dblp.uni-trier.de/rec/bibtex/#1.xml}
  {dblp:#1}}
\def\mn@eprint@#1:#2:#3:#4\@nil{\def\@tempa {#1}\def\@tempb {#2}\def\@tempc
  {#3}\ifx \@tempc \@empty \let \@tempc \@tempb \let \@tempb \@tempa \fi \ifx
  \@tempb \@empty \def\@tempb {arXiv}\fi \@ifundefined
  {mn@eprint@\@tempb}{\@tempb:\@tempc}{\expandafter \expandafter \csname
  mn@eprint@\@tempb\endcsname \expandafter{\@tempc}}}

\bibitem[\protect\citeauthoryear{Beasor \& Davies}{Beasor \&
  Davies}{2016}]{beasor2016evolution}
Beasor E.~R.,  Davies B.,  2016, Monthly Notices of the Royal Astronomical
  Society, 463, 1269

\bibitem[\protect\citeauthoryear{Bonanos et~al.,}{Bonanos
  et~al.}{2010}]{bonanos2010spitzer}
Bonanos A.,  et~al., 2010, The Astronomical Journal, 140, 416

\bibitem[\protect\citeauthoryear{Currie et~al.,}{Currie
  et~al.}{2010}]{currie2010stellar}
Currie T.,  et~al., 2010, The Astrophysical Journal Supplement Series, 186, 191

\bibitem[\protect\citeauthoryear{Davies \& Beasor}{Davies \&
  Beasor}{2017}]{davies2017initial}
Davies B.,  Beasor E.~R.,  2017, \url {https://arxiv.org/abs/1709.06116}

\bibitem[\protect\citeauthoryear{De~Beck, Decin, de Koter, Justtanont,
  Verhoelst, Kemper  \& Menten}{De~Beck et~al.}{2010}]{de2010probing}
De~Beck E.,  Decin L.,  de Koter A.,  Justtanont K.,  Verhoelst T.,  Kemper F.,
    Menten K.~M.,  2010, Astronomy \& Astrophysics, 523, A18

\bibitem[\protect\citeauthoryear{De~Jager, Nieuwenhuijzen  \& Van
  Der~Hucht}{De~Jager et~al.}{1988}]{de1988mass}
De~Jager C.,  Nieuwenhuijzen H.,   Van Der~Hucht K.,  1988, Astronomy and
  Astrophysics Supplement Series, 72, 259

\bibitem[\protect\citeauthoryear{Dotter}{Dotter}{2016}]{dotter2016mesa}
Dotter A.,  2016, The Astrophysical Journal Supplement Series, 222, 8

\bibitem[\protect\citeauthoryear{Draine \& Lee}{Draine \&
  Lee}{1984}]{draine1984optical}
Draine B.,  Lee H.~M.,  1984, The Astrophysical Journal, 285, 89

\bibitem[\protect\citeauthoryear{Ekstr{\"o}m et~al.,}{Ekstr{\"o}m
  et~al.}{2012}]{ekstrom2012grids}
Ekstr{\"o}m S.,  et~al., 2012, Astronomy \& Astrophysics, 537, A146

\bibitem[\protect\citeauthoryear{Eldridge \& Stanway}{Eldridge \&
  Stanway}{2009}]{eldridge2009spectral}
Eldridge J.~J.,  Stanway E.~R.,  2009, Monthly Notices of the Royal
  Astronomical Society, 400, 1019

\bibitem[\protect\citeauthoryear{Feast \& Whitelock}{Feast \&
  Whitelock}{1992}]{feast1992ch}
Feast M.,  Whitelock P.,  1992, Monthly Notices of the Royal Astronomical
  Society, 259, 6

\bibitem[\protect\citeauthoryear{Gazak, Davies, Kudritzki, Bergemann  \&
  Plez}{Gazak et~al.}{2014}]{gazak2014quantitative}
Gazak J.~Z.,  Davies B.,  Kudritzki R.,  Bergemann M.,   Plez B.,  2014, The
  Astrophysical Journal, 788, 58

\bibitem[\protect\citeauthoryear{Georgy}{Georgy}{2012}]{georgy2012yellow}
Georgy C.,  2012, Astronomy \& Astrophysics, 538, L8

\bibitem[\protect\citeauthoryear{Georgy \& Ekstr{\"o}m}{Georgy \&
  Ekstr{\"o}m}{2015}]{georgy2015mass}
Georgy C.,  Ekstr{\"o}m S.,  2015, arXiv preprint arXiv:1508.04656

\bibitem[\protect\citeauthoryear{Georgy et~al.,}{Georgy
  et~al.}{2013}]{georgy2013grids}
Georgy C.,  et~al., 2013, Astronomy \& Astrophysics, 558, A103

\bibitem[\protect\citeauthoryear{Goldman et~al.,}{Goldman
  et~al.}{2017}]{goldman2017wind}
Goldman S.~R.,  et~al., 2017, Monthly Notices of the Royal Astronomical
  Society, 465, 403

\bibitem[\protect\citeauthoryear{Gontcharov}{Gontcharov}{2016}]{gontcharov2016extinction}
Gontcharov G.,  2016, Astronomy Letters, 42, 445

\bibitem[\protect\citeauthoryear{Ivezic, Nenkova  \& Elitzur}{Ivezic
  et~al.}{1999}]{ivezic1999dusty}
Ivezic Z.,  Nenkova M.,   Elitzur M.,  1999, Astrophysics Source Code Library,
  1, 11001

\bibitem[\protect\citeauthoryear{Koornneef}{Koornneef}{1983}]{koornneef1983near}
Koornneef J.,  1983, Astronomy and Astrophysics, 128, 84

\bibitem[\protect\citeauthoryear{Levesque, Massey, Olsen, Plez, Maeder  \&
  Meynet}{Levesque et~al.}{2005}]{levesque2005physical}
Levesque E.,  Massey P.,  Olsen K.,  Plez B.,  Maeder A.,   Meynet G.,  2005,
  in Bulletin of the American Astronomical Society. p.~1465

\bibitem[\protect\citeauthoryear{Marco \& Negueruela}{Marco \&
  Negueruela}{2013}]{marco2013ngc}
Marco A.,  Negueruela I.,  2013, Astronomy \& Astrophysics, 552, A92

\bibitem[\protect\citeauthoryear{Maund et~al.,}{Maund
  et~al.}{2011}]{maund2011yellow}
Maund J.~R.,  et~al., 2011, The Astrophysical Journal Letters, 739, L37

\bibitem[\protect\citeauthoryear{Mauron \& Josselin}{Mauron \&
  Josselin}{2011}]{mauron2011mass}
Mauron N.,  Josselin E.,  2011, Astronomy \& Astrophysics, 526, A156

\bibitem[\protect\citeauthoryear{Messineo, Habing, Menten, Omont, Sjouwerman
  \& Bertoldi}{Messineo et~al.}{2005}]{messineo200586}
Messineo M.,  Habing H.,  Menten K.,  Omont A.,  Sjouwerman L.,   Bertoldi F.,
  2005, Astronomy \& Astrophysics, 435, 575

\bibitem[\protect\citeauthoryear{Morozova, Piro  \& Valenti}{Morozova
  et~al.}{2017}]{morozova2017unifying}
Morozova V.,  Piro A.~L.,   Valenti S.,  2017, The Astrophysical Journal, 838,
  28

\bibitem[\protect\citeauthoryear{Nieuwenhuijzen \& De~Jager}{Nieuwenhuijzen \&
  De~Jager}{1990}]{nieuwenhuijzen1990parametrization}
Nieuwenhuijzen H.,  De~Jager C.,  1990, Astronomy and Astrophysics, 231, 134

\bibitem[\protect\citeauthoryear{Price, Egan, Carey, Mizuno  \& Kuchar}{Price
  et~al.}{2001}]{price2001midcourse}
Price S.~D.,  Egan M.~P.,  Carey S.~J.,  Mizuno D.~R.,   Kuchar T.~A.,  2001,
  The Astronomical Journal, 121, 2819

\bibitem[\protect\citeauthoryear{Reimers}{Reimers}{1975}]{reimers1975circumstellar}
Reimers D.,  1975, Memoires of the Societe Royale des Sciences de Liege, 8, 369

\bibitem[\protect\citeauthoryear{Richards \& Yates}{Richards \&
  Yates}{1998}]{richards1998maser}
Richards A.,  Yates J.,  1998, Irish Astronomical Journal, 25, 7

\bibitem[\protect\citeauthoryear{Scicluna, Siebenmorgen, Wesson, Blommaert,
  Kasper, Voshchinnikov  \& Wolf}{Scicluna et~al.}{2015}]{scicluna2015large}
Scicluna P.,  Siebenmorgen R.,  Wesson R.,  Blommaert J.,  Kasper M.,
  Voshchinnikov N.,   Wolf S.,  2015, Online Material p, 1

\bibitem[\protect\citeauthoryear{Shenoy et~al.,}{Shenoy
  et~al.}{2016}]{shenoy2016searching}
Shenoy D.,  et~al., 2016, The Astronomical Journal, 151, 51

\bibitem[\protect\citeauthoryear{Skrutskie et~al.,}{Skrutskie
  et~al.}{2006}]{skrutskie2006two}
Skrutskie M.,  et~al., 2006, The Astronomical Journal, 131, 1163

\bibitem[\protect\citeauthoryear{Smartt}{Smartt}{2015}]{smartt2015observational}
Smartt S.,  2015, Publications of the Astronomical Society of Australia, 32,
  e016

\bibitem[\protect\citeauthoryear{Smartt, Eldridge, Crockett  \& Maund}{Smartt
  et~al.}{2009}]{smartt2009death}
Smartt S.,  Eldridge J.,  Crockett R.,   Maund J.~R.,  2009, Monthly Notices of
  the Royal Astronomical Society, 395, 1409

\bibitem[\protect\citeauthoryear{Smith}{Smith}{2014}]{smith2014mass}
Smith N.,  2014, Annual Review of Astronomy and Astrophysics, 52, 487

\bibitem[\protect\citeauthoryear{Smith, Humphreys, Davidson, Gehrz, Schuster
  \& Krautter}{Smith et~al.}{2001}]{smith2001asymmetric}
Smith N.,  Humphreys R.~M.,  Davidson K.,  Gehrz R.~D.,  Schuster M.,
  Krautter J.,  2001, The Astronomical Journal, 121, 1111

\bibitem[\protect\citeauthoryear{Smith, Hinkle  \& Ryde}{Smith
  et~al.}{2009}]{smith2009red}
Smith N.,  Hinkle K.~H.,   Ryde N.,  2009, The Astronomical Journal, 137, 3558

\bibitem[\protect\citeauthoryear{Smith, Li, Filippenko  \& Chornock}{Smith
  et~al.}{2011}]{smith2011observed}
Smith N.,  Li W.,  Filippenko A.~V.,   Chornock R.,  2011, Monthly Notices of
  the Royal Astronomical Society, 412, 1522

\bibitem[\protect\citeauthoryear{Smith et~al.,}{Smith
  et~al.}{2015}]{smith2015PTF11iqb}
Smith N.,  et~al., 2015, Monthly Notices of the Royal Astronomical Society,
  449, 1876

\bibitem[\protect\citeauthoryear{Smith et~al.,}{Smith
  et~al.}{2016}]{smith2017endurance}
Smith N.,  et~al., 2016, Monthly Notices of the Royal Astronomical Society, p.
  stw3204

\bibitem[\protect\citeauthoryear{Van~Loon, Zijlstra, Bujarrabal  \&
  Nyman}{Van~Loon et~al.}{2001}]{van2001circumstellar}
Van~Loon J.~T.,  Zijlstra A.~A.,  Bujarrabal V.,   Nyman L.-{\AA}.,  2001,
  Astronomy \& Astrophysics, 368, 950

\bibitem[\protect\citeauthoryear{Van~Loon, Cioni, Zijlstra  \& Loup}{Van~Loon
  et~al.}{2005}]{van2005empirical}
Van~Loon J.~T.,  Cioni M.-R.,  Zijlstra A.~A.,   Loup C.,  2005, Astronomy \&
  Astrophysics, 438, 273

\bibitem[\protect\citeauthoryear{Wright et~al.,}{Wright
  et~al.}{2010}]{wright2010wide}
Wright E.~L.,  et~al., 2010, The Astronomical Journal, 140, 1868

\makeatother
\end{thebibliography}



\appendix

\section{Some extra material}

If you want to present additional material which would interrupt the flow of the main paper,
it can be placed in an Appendix which appears after the list of references.


\bsp	
\label{lastpage}
\end{document}